\documentclass[twocolumn,prd,showpacs,10pt,superscriptaddress,nofootinbib]{revtex4}
\usepackage{graphics}

\def\be{\begin{equation}}
\def\ee{\end{equation}}
\def\bea{\begin{eqnarray}}
\def\eea{\end{eqnarray}}
\def\ba{\begin{array}}
\def\ea{\end{array}}
\newcommand{\lsim}{\,\raise 0.4ex\hbox{$<$}\kern -0.8em\lower 0.62ex\hbox{$\sim$}\,}
\newcommand{\gsim}{\,\raise 0.4ex\hbox{$>$}\kern -0.7em\lower 0.62ex\hbox{$\sim$}\,}
\newcommand{\mpl}{{M_\mathrm{Pl}^2}}

\begin{document}

\title{A dark energy view of inflation}
\date{April 7, 2010}

\author{St\'ephane Ili\'c}
\email{stephane.ilic@u-psud.fr}
\affiliation{Magist\`ere de Physique
Fondamentale, Universit\'e Paris-Sud XI, Orsay 91405, France}
\affiliation{Astronomy Centre, University of Sussex, Falmer, Brighton
BN1 9QH, UK}
\author{Martin Kunz}
\email{Martin.Kunz@unige.ch}
\affiliation{D\'epartement de Physique Th\'eorique, Universit\'e de Gen\`eve, 1211 Geneva 4, Switzerland}
\affiliation{Astronomy Centre, University of Sussex, Falmer, Brighton
BN1 9QH, UK}
\author{Andrew R. Liddle}
\email{A.Liddle@sussex.ac.uk}
\affiliation{Astronomy Centre, University of Sussex, Falmer, Brighton
BN1 9QH, UK}
\author{Joshua A. Frieman}
\email{frieman@fnal.gov}
\affiliation{Fermilab Center for Particle Astrophysics, Batavia, IL 60510, USA}
\affiliation{Kavli Institute for Cosmological Physics, The University of Chicago, Chicago, IL 60637, USA}

\begin{abstract}
Traditionally, inflationary models are analyzed in terms of parameters
such as the scalar spectral  index $n_s$ and the tensor to scalar ratio $r$,
while dark energy models are studied in terms of the equation of state
parameter $w$. Motivated by the fact that both deal with periods of
accelerated expansion, we study the evolution of $w$ during
inflation, in order to derive observational constraints on its value during an earlier
epoch likely dominated by a dynamic form of dark energy. We find that
the cosmic microwave background and large-scale structure data is 
consistent with $w_{\rm inflation}=-1$
and provides an upper limit of $1+w\lsim 0.02$. Nonetheless, an exact de Sitter 
expansion with a constant $w=-1$ is disfavored since this
would result in $n_s=1$.
\end{abstract}

\maketitle

\section{Introduction}

The nature of the dark energy has been seen as one of the principal puzzles in cosmology, and
in theoretical physics as a whole, ever since the supernova observations
\cite{sn1,sn2} in 1998 confirmed the mounting suspicion that the expansion
rate of the Universe is accelerating. One of the leading contenders is
the cosmological constant, for which the equation of state $w$ equals $-1$, both on theoretical grounds
and because no confirmed deviations from $w=-1$ have come from cosmological
observations.

However, the current phase of accelerated expansion is most likely not
the only one in the history of the Universe: it is thought that a 
much earlier epoch of accelerated expansion called inflation created 
the initial fluctuations that led to large-scale structure and solved several 
problems of the standard Big Bang cosmology. The spectrum of fluctuations that we observe today, particularly
in the cosmic microwave background  (CMB) radiation, indicates that they were
created by a mechanism that was able to act outside the normal causal
horizon \cite{zalda,skd}. It is commonly believed that the structure we see in the 
CMB and in the distribution of galaxies arose from quantum fluctuations 
that were stretched outside the Hubble horizon by a phase
of accelerated expansion, not dissimilar to the one that is being
observed today.

We know that inflation ended early in cosmic history, before 
the epoch of Big Bang nucleosynthesis: an inflating Universe is nearly 
empty of matter and does not form galaxies. As a consequence, inflation 
could not have been driven by a pure cosmological constant. Since the Universe 
apparently began to inflate again several billion years ago, it is natural to ask 
whether hypothetical observers present during primordial inflation would have 
been able to distinguish between a cosmological constant and an alternative 
model such as a scalar field by studying the expansion history quantified by 
$w$. In this paper, we will link the usual inflationary observables
to $w$ and provide constraints on $w$ during the period when the observable
scales left the horizon.


\section{The equation of state of the inflaton\label{sec:eos}}

We assume that inflation started well before the observable scales left the
horizon, i.e., that it lasted longer than about 60 e-folds of 
expansion, so that the only significant contribution to the energy density $\rho$ is
the one from the inflaton itself and that the Universe can be taken to be
spatially flat. This implies that the Friedmann and energy
conservation equations are
\bea
H^2 &=& \frac{\rho}{3 \mpl} \, , \\
\dot{\rho} &=& -3 H(1+w)\rho \, .
\eea
Here we used the reduced Planck mass, $\mpl \equiv 1/ 8\pi G$ in our units
where $c=\hbar=1$,
and the Hubble parameter $H\equiv\dot{a}/a$ where $a$ is the scale factor.
We can compute the equation of state parameter $w=p/\rho$ once
we know the expansion rate $H$,
\be
1+w = -\frac{2}{3}\frac{\dot{H}}{H^2} \, . \label{eq:wH}
\ee
It is of course equally possible to compute $w$ directly from
the pressure and the energy density of the inflaton. However, the form
given above is especially useful in the case of single-field inflation,
in which case the perturbations generated are linked to
$H$ as there is only a single degree of freedom present (exemplified by the potential of the inflaton field). This allows us to connect the expression for $w$
directly to quantities related to the perturbations.

This turns out to be especially simple when working with the slow-roll
parameters in the so-called Hamilton--Jacobi formalism, see
e.g.~Ref.~\cite{infla_review} 
for detailed derivations. The first two slow-roll parameters are defined as
\bea
\epsilon_H &=& 2 \mpl \left(\frac{H'}{H}\right)^2 , \\
\eta_H &=& 2 \mpl \frac{H''}{H} .
\eea
Here $'$ denotes a derivative with respect to the scalar field $\phi$. Since
$H' = \dot{H}/\dot{\phi}$ and $\dot{\phi} = -2 \mpl H'$, we find together
with Eq.~(\ref{eq:wH}) that
\be
1+w = \frac{2}{3} \epsilon_H .
\ee
The equation of state during inflation is therefore directly given by
the first slow-roll parameter. To lowest order in slow-roll this is
also related to the tensor to scalar ratio by $r = 16\epsilon_H$. Without
any further work we can deduce that, since primordial gravitational waves 
have not been observed, there is no observational requirement for a 
deviation from $w=-1$ during inflation. The upper limit on $r$
from the five-year Wilkinson Microwave Anisotropy Probe (WMAP)
data for a flat
$\Lambda$CDM model without running is about $0.43$ \cite{wmap5},  corresponding to
a maximum deviation 
from $w=-1$ of $0.02$. We will derive precise numerical constraints in the next
section.\footnote{After we completed the calculations for this paper, the WMAP
team released the 7-year data (WMAP7). It gives results very similar to WMAP5 and we 
would not expect qualitative, or even significant quantitative, changes. For example,
the upper limit on $r$ decreases only slightly to $0.37$.}

This result is at first glance a bit puzzling: An equation of state
$w=-1$ leads to de Sitter expansion which in turn creates a scale-invariant
Harrison--Zel'dovich (HZ) spectrum. However, the WMAP five-year data paper
also claims a 2.5 sigma
deviation from a HZ spectrum. 
The explanation is that the deviation of the
scalar spectral index $n_s$ from the HZ case ($n_s=1$) can be caused by the
second slow-roll parameter $\eta_H$, given to lowest order in slow roll by
\be
2\eta_H = \left(n_s-1+4\epsilon_H\right) .
\label{eq:ns}
\ee
Thus even if at a given time $\epsilon_H \approx 0$, it is still possible
to obtain $n_s\neq 1$ through a non-zero $\eta_H$.

A non-zero $\eta_H$ implies that $w$ will evolve away from $-1$.
How quickly will it do that? Possibly fast enough to lead to measurable
deviations during the observable number of $e$-foldings? We find
\be
\frac{d\ln (1+w)}{dN} = \frac{d\ln \epsilon_H}{dN} = 2 (\eta_H-\epsilon_H)
\label{eq:wconst}
\ee
where $N=-\ln a$ is the number of $e$-foldings. 
Since the rate of change
of $\epsilon_H$ is proportional to $\epsilon_H$ itself, it can become very small if
$\epsilon_H$ is very small. Close to de Sitter the field freezes and
moves only very slowly, but even this slow motion leads to observable
effects in the power spectrum of the perturbations. This is unfortunately
an observational channel that is not available for the contemporary
dark energy. Indeed, the ways in which we probe inflation and today's
dark energy are very different: we have no way to constrain directly
the expansion history during inflation, but we can see the spectrum of
the curvature perturbations generated during this epoch. On the other
hand, while we can observe directly the recent expansion history of the
Universe and infer the equation of the state of the dark energy, the
fluctuations generated during the current bout of accelerated expansion
are impossible to observe both because of their tiny predicted amplitude 
and because they become classical only when outside the current horizon.

The likelihood of a tiny value of $\epsilon_H$ has been hotly debated in the inflation literature (e.g.~Refs.~\cite{lyth,george}), since it would prevent direct detection of inflationary gravitational waves, e.g.\ by a CMB polarization satellite mission \cite{cmbpol}. Within the framework of the early large-field inflation models, such as monomial potentials, a tiny $\epsilon_H$ and large $\eta_H$ would look rather unnatural, and hence the observed
$n_{\rm s} \approx 0.96$ would suggest $r\approx 0.1-0.2$ and
$1+w\approx0.01$, both well within current experimental bounds. However by contrast the paradigm of small-field models, such as hybrid inflation, motivated by the need to keep the field values small in a supergravity context, does suggest that $\epsilon_H$ must be extremely small at horizon crossing, thus indicating $w$ very close to $-1$.

\section{Numerical investigation}

In order to obtain numerical constraints on $w$ during inflation, we need
to link it to observational quantities. In this paper we will use the spectrum of
the primordial fluctuations, as observed in the CMB. The link between $w$
and $H$ given in Eq.~(\ref{eq:wH}) is
fundamental, failing to hold only if either the universe was very
different from Friedmannian during inflation or if there were other
contributions to the expansion rate present. The first would invalidate
the whole inflationary framework, while in the second case our $w$ would
correspond to an effective total $w$.

To go from $H$ to the primordial power spectrum requires a specific
model. Here we assume that inflation was due to a single canonical
scalar field, though without making the common assumption of
slow-roll. An interesting future project is to relax this condition by
investigating a range of other models, for example K-inflation models
with a different sound speed \cite{kinf}. While this may change quantitative
limits on $w$, we do not expect it to change the qualitative results.
We also note that our model imposes the Dominant Energy Condition 
$w\geq -1$ by construction. 

In order to compute $H$ during the observable range of scales, we use the 
module provided by Lesgourgues and Valkenburg (LV) \cite{lv} which takes the slow-roll 
parameters at the pivot scale as an input.
The pivot scale is fixed to be $k_* = 0.01$/Mpc (roughly in the
center of the observable range); this is the scale
at which the Hubble parameter $H$ is expanded as a Taylor series
in $(\phi-\phi_*)$, with the scale factor set to $a_* = k_*/H_*$.
We then approach the problem from two slightly different angles. To reconstruct
the evolution of $w$ from the time when the observable scales left the
horizon up to the end of inflation, we use
the flow-equation formalism \cite{flow} to derive the evolution of
$\epsilon_H$ from the end of inflation to the observable scales. In
this we proceed similarly to Ref.~\cite{hk} by selecting  ``initial"
values for the first four slow-roll parameters at
the end of inflation (in fact we only
choose three of them since $\epsilon_H$ is always equal to 1) 
and flowing them back 60 $e$-foldings using the
flow equations. The values obtained at $N=60$ are then used to
compute the observables using the LV module. The appropriate value
of $N$ changes with the inflationary energy scale, and may be
smaller in low energy scales models. But this does not impact
our conclusions, since we are interested in
the experimental constraints on $w$ around the scales that are
directly probed by observations, and additionally, as our later results show, the constraints
remain fairly constant over a range of $N$. This allows us to avoid more sophisticated approaches 
to treating this uncertainty, as given for instance in Ref.~\cite{pe08}.

On the other hand, we do not really know what happened after the observable scales left
the horizon, as we do not have any observations concerning  that
period. Based on this reasoning, Lesgourgues and Valkenburg \cite{lv} 
argued that considering only the observable scales makes it
possible to work with a relatively low-order expansion of the scalar
field potential without introducing artificial constraints. We use the
module provided by LV to compute the observables in their framework
and to compare the results with those from the flow-equation
formalism.

\begin{figure}
\scalebox{0.7}{\includegraphics{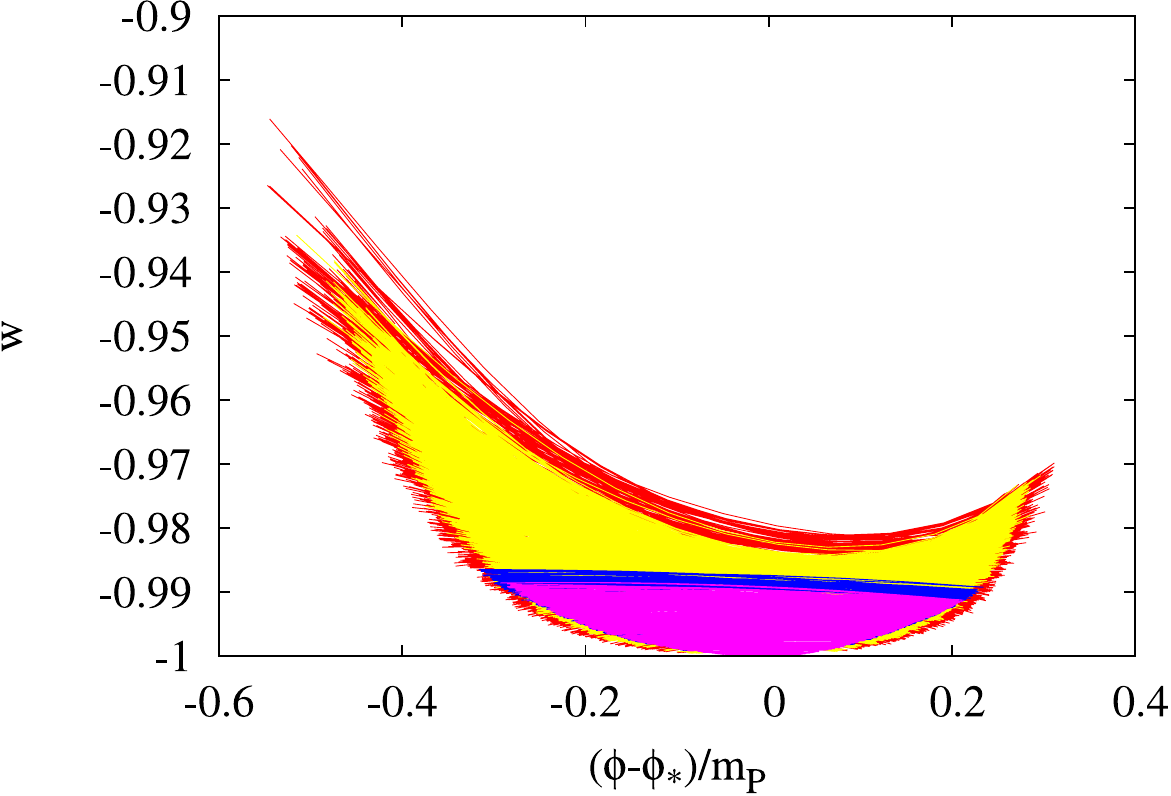}}
\caption{The evolution of $w(\phi)$ for a range of cases accepted by the CMB likelihood,
around the field value $\phi_*$ at the pivot scale $k_* = 0.01$/Mpc. 
The red and yellow curves lie within the 95\% and 68\% confidence regions for the LV formalism, 
blue and purple give the same information for the flow-equation formalism. From the outside inward, the colored regions are red, yellow, blue, and purple.
}
\label{fig:w_phi}
\end{figure}

In both cases we use CosmoMC \cite{cosmomc} to perform a Markov Chain
Monte Carlo exploration of the parameter space, which includes, depending on our method:
\begin{itemize}
\item the slow-roll parameters at $N=0$, i.e. at the end of inflation, for our first approch; the useful parameters at the pivot scale (fixed at $N=60$) are computed by solving numerically the flow equations;
\item the slow-roll parameters at the pivot scale directly for the second method.
\end{itemize}
Here we used the first four slow-roll parameters $\epsilon_H$, $\eta_H$, $^2\beta_H=\xi_H$ and $^3\beta_H$ \cite{floweq}
 with the following ranges (at $N=0$ for the first method, and at the pivot scale for the second):  $\epsilon_H \in [0.0,1.0]$ (fixed to 1.0 for the first method), and $\eta_H$, $\xi_H$, $^3\beta_H$ $\in [-10.0,10.0]$. 
CosmoMC works together with CAMB \cite{camb} to compute the CMB power spectrum and
then uses the WMAP five-year likelihood code \cite{wmap5}. The inflationary power spectrum is calculated using the Lesgourgues--Valkenburg module which solves the perturbation mode equation. This setup allows us to compute chains of
acceptable expansion histories during inflation. These were then
mapped into chains of $w(\phi)$ (Fig.~\ref{fig:w_phi}).

\begin{figure}
\scalebox{0.8}{\includegraphics{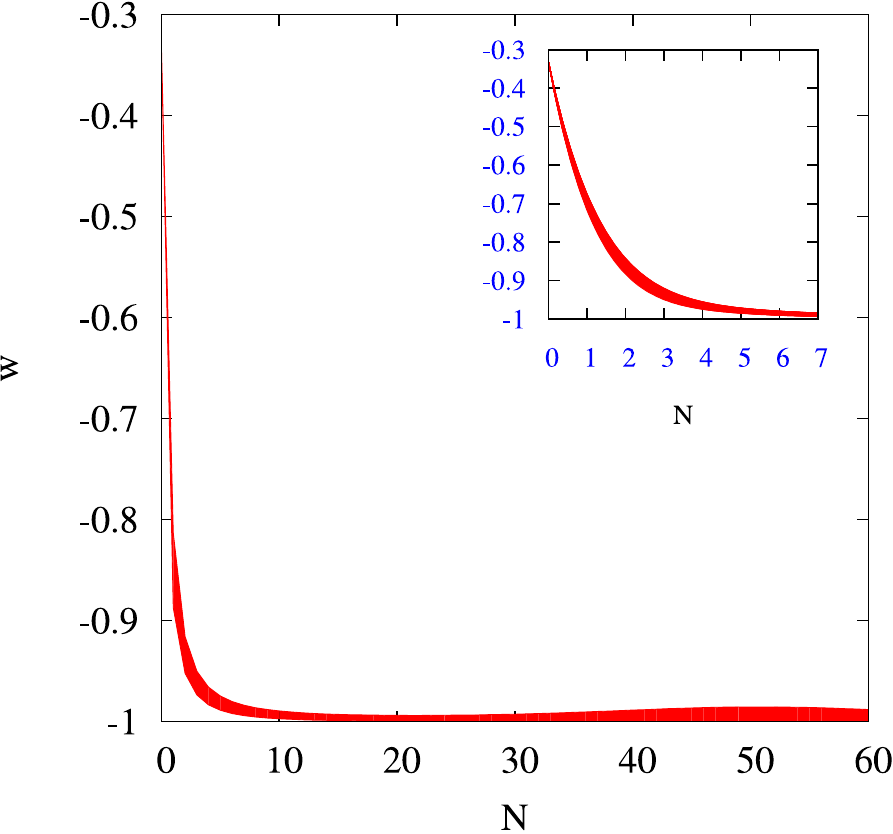}}
\caption{The complete evolution of $w(N)$, from the flow-equation results accepted by the CMB likelihood. Inflation is made to end at $N=0$ where $w(N=0)=-1/3$ corresponding to $\epsilon_H(N=0)=1$. For our choice of
priors on the slow-roll parameters at $N=0$, we find that $w$ decreases
rapidly towards $-1$ (see inset) and stays close to it during the period when the observable
scales leave the horizon ($N\approx 40 - 60$).}
\label{fig:w_N}
\end{figure}

\begin{figure}
\scalebox{0.7}{\includegraphics{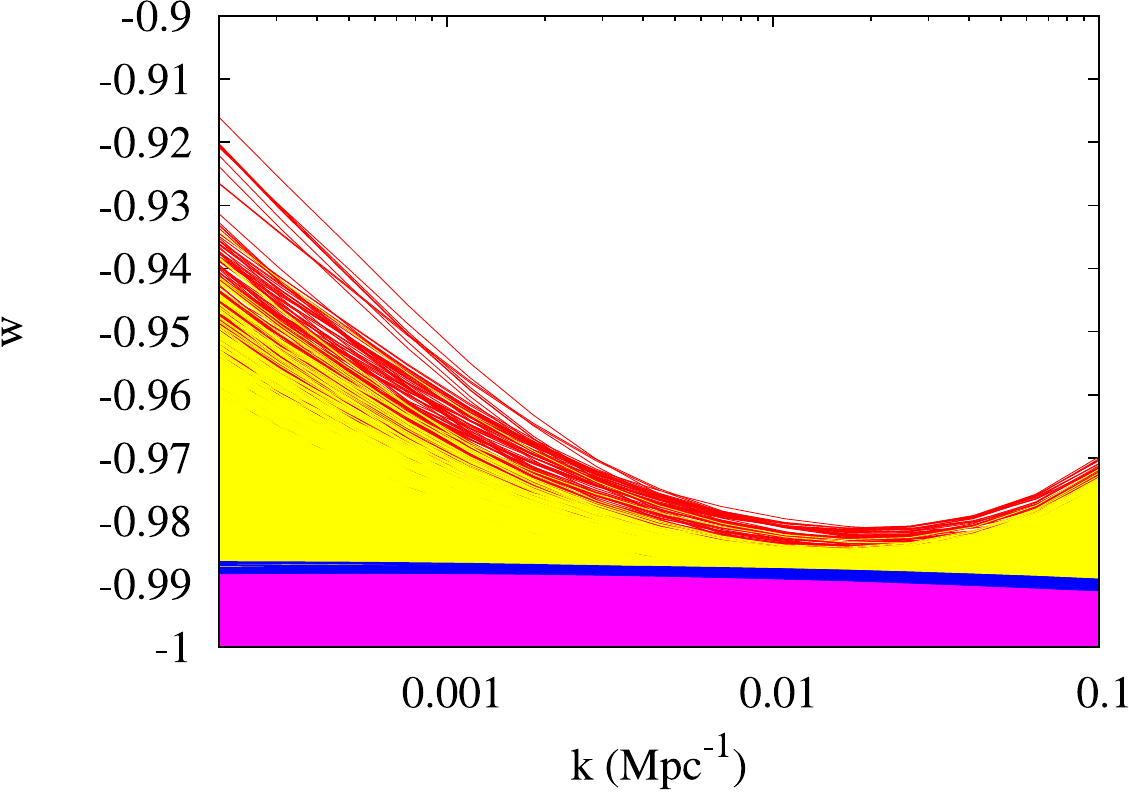}}
\caption{The evolution of $w$ as a function of the comoving scale $k$, using only the 5-year WMAP CMB data. Red and yellow are the 95\% and 68\% confidence regions for the LV formalism. Blue and purple are the same for the flow-equation formalism.
From the outside inward, the colored regions are red, yellow, blue, and purple.
}
\label{fig:w_k}
\end{figure}

In principle Fig.~\ref{fig:w_phi} already shows the constraints on the
equation of state parameter during inflation. But, as is easily seen
in the figure, $\phi$ moves more and more slowly as we approach $w=-1$,
which makes the constraints difficult to interpret. A better representation is
$w(N)$ in terms of the number of $e$-foldings $N$ before the end of
inflation, see Fig.~\ref{fig:w_N}. However, in the LV formalism the
field is never evolved until the end of inflation, so that $N$ is not
defined.  An alternative way to plot the results in this situation is
to map them instead to the horizon scale at that epoch, $k=aH$. Since
the perturbations freeze in outside the horizon and turn into
conserved curvature perturbations, this scale corresponds to the one
that they have when they re-enter the horizon. We plot our constraints
in this way in Fig.~\ref{fig:w_k}.

>From the full evolution in Fig.~\ref{fig:w_N} we see that $w$
approaches $-1$ rapidly as we move into the past. The precise rate at
which $-1$ is approached depends on the range of models chosen at the
end of inflation (see e.g.~Ref.~\cite{ekp06}). Nonetheless, as shown in the inset, strong deviations
from $w=-1$ are expected in the last few e-folds. To illustrate the
scales involved: if we were to arbitrarily place today at $N=7$
(the right-hand limit of the inset) and reverse time then $N=0$ would
roughly correspond to last scattering ($z\approx 1100$).

The current experimental uncertainty on the dark energy $w$ is about $0.1$, comfortably enclosing $w=-1$, and in the
future will reach a precision of $0.02$ or better. We find 
that the current limits on $w$ during inflation are comparable, with
a 95\% limit of $1+w < 0.02$ at $k\approx 0.01$/Mpc, see Fig.~\ref{fig:w_k}.
This agrees well with the arguments in the previous section, but
the figure shows also the precise shape of the constraints.
There is no lower limit on $w$ (apart from $w \geq -1$ enforced by the 
model construction). However, the tentative observation of a deviation from a
scale-invariant primordial power spectrum implies through Eq.~(\ref{eq:ns})
that $\epsilon_H$ and $\eta_H$ cannot both be zero. Together
with Eq.~(\ref{eq:wconst}) this disfavours a constant $w=-1$. But as
discussed in Section \ref{sec:eos}, $(1+w)$ can remain small over
the observable range of scales and we find that this deviation
is not visible in the figures.

\begin{figure}
\scalebox{0.7}{\includegraphics{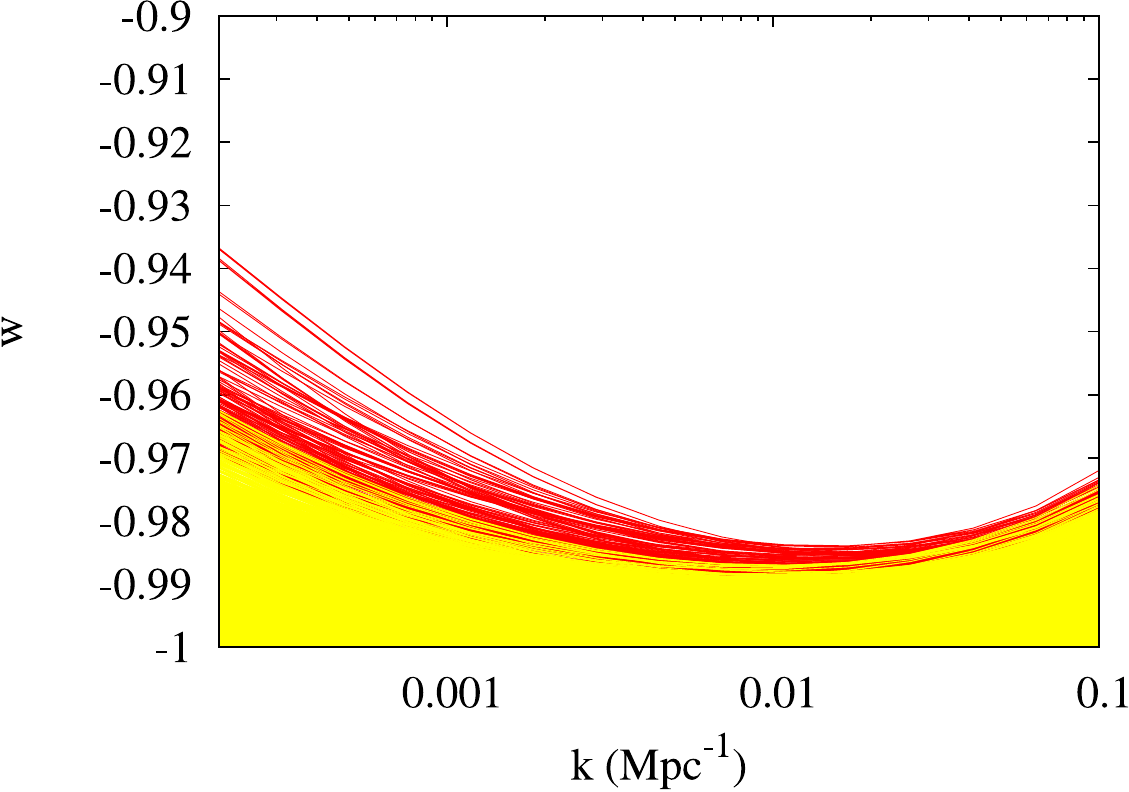}}
\caption{The evolution of $w$ as a function of the comoving scale $k$, as
in Fig.~\ref{fig:w_k}, but using in addition the SDSS DR7 LRG data. We only
show the LV limits.}
\label{fig:cmbsdss}
\end{figure}

The limits on $w$ can be improved by extending the lever arm
of the measurements, for example by adding galaxy survey data
on smaller scales. We show the impact of using both WMAP 5-year CMB data
and Sloan Digital Sky Survey (SDSS) Data Release 7 Luminous Red Galaxy
data (DR7 LRG) \cite{sdss} in Fig.~\ref{fig:cmbsdss}. The shape of the constraints
have not changed by much, but the limits have become somewhat tighter.
We can achieve another small increase in precision by adding further CMB data
on smaller scales, but again the improvement is small so we do
not show those constraints.

We also notice that the prescription of LV allows for a stronger
variation of $w$. The flow-equation formalism with the number
of parameters and priors used here leads to very little evolution
of $w$ during the observable period. This does not mean that
one of the two approaches is wrong, but rather that they impose
different additional conditions. As always, it is important to be
aware of these effective (and somewhat hidden) priors.

\section{Conclusions}

It seems very likely that there have been at least two periods of
accelerated expansion during the evolution of the Universe. During the
first period, called inflation, the perturbations that led to today's
structure were generated, while the second one has started only
recently and is attributed to a mysterious dark energy. In this paper
we ask what a similar physical origin would imply for the 
dark energy.

One point that is immediately clear is that since inflation ended,
there is reason to assume that it was not due to a cosmological
constant.  This is supported by the tentative detection of a deviation
from an exact Harrison-Zel'dovich spectrum with $n_s=1$ \cite{wmap5}:
a period dominated by a (possibly effective) cosmological constant
would either result in no perturbations at all or in perturbations with an
exactly scale-invariant spectrum, depending on how precisely the de Sitter
state is reached. The former possibility is clearly excluded, and while current
observations are not yet conclusive on whether $n_s=1$
is excluded, the Planck
satellite should settle the question within the next few years, since it
is expected to reach a precision of $\sigma_{n_s}\lesssim 0.005$
\cite{ns_planck}. A clear detection of $n_s\neq 1$ would require
either $w'\neq0$ or $w\neq-1$, with a constant $w=-1$ being ruled
out in both cases.

However, there is also no requirement for the equation of state
parameter $w$ to differ appreciably from $-1$ during inflation as $w$
is directly proportional to the ratio of tensor to scalar
perturbations, and no primordial gravitational waves have been
detected so far. Thus, even though it may be possible that Planck
demonstrates that inflation was {\em not} due to a cosmological
constant, this does not imply that $w$ was measurably different from
$-1$. Indeed, we find that current data allows $w$ to be arbitrarily
close to $-1$ as long as it changes just slightly during its
evolution. This direct link between $w$ and the gravitational wave
background reinforces the importance of the latter as a probe of early
Universe physics: if it is detected then we know immediately that $w$
was measurably different from $-1$ during inflation.

We have also found that the current experimental limits on $w$ during
inflation imply $1+w<0.02$ at a scale of $k\approx 0.01$/Mpc.  If we
take seriously the idea that early and late-time acceleration are
based on similar mechanisms, then this might suggest that dark energy probes
need to reach at least this precision in order to have a reasonable chance
of detecting any deviation from $\Lambda$. Following the arguments
from the end of section II, one could argue for a target precision of
about $0.01$ for measuring $w$, beyond which there may well be a ``$w$ desert''
extending to very low values of $(1+w)$. This precision also roughly leads to 
a decisive Bayes factor in favour of $\Lambda$CDM if no deviation 
from $-1$ is detected (when looking at constant $w$, see e.g.~Ref.~\cite{pia}
for the methodology).

However, the absence of an observational lower limit on $w$ during inflation
should not be taken as argument against measuring the
recent expansion history and evolution of perturbations. Firstly,
there is no direct evidence that the two periods of accelerated
expansion are due to the same underlying physical mechanism. Secondly,
even if that is so, it is likely that we are observing a very
different epoch of the inflationary phenomenon today than in the early
Universe.  The acceleration became observationally relevant only very
recently, less than one $e$-folding ago. If the onset of acceleration
coincides with it becoming visible, then we could expect strong
deviations from $w=-1$, since also at the end of inflation $w$
deviated strongly from $-1$, see Fig.~\ref{fig:w_N}. On the other
hand, it is also possible that the dark energy has been present much
longer but was buried beneath the matter and has but surfaced
recently. In this case inflation indicates that it is natural for a
scalar field dark energy to have an equation of state close to
$p=-\rho$.

Finally, inflation and the current epoch are accessible in very
different ways: from inflation we observe the curvature
perturbations generated out of quantum fluctuations, while for the
recent history of the Universe we instead observe directly the
evolution of the expansion history as well as possibly the impact of
the dark energy perturbation or of deviations from General Relativity
onto light deflection and the distribution of galaxies. If the physics underlying
the accelerated expansion of inflation and dark energy are related, then
the two sets of observations are complementary and mutually reinforcing,
and observational results for either period of accelerated expansion
may help to shed light on the other one as well.

\begin{acknowledgments}

M.K. and A.R.L. are supported by STFC (UK). It is a pleasure to thank
Chiara Caprini for helpful and interesting discussions.

\end{acknowledgments}

\end{document}